\documentclass[manuscript, screen]{acmart}
\AtBeginDocument{%
  \providecommand\BibTeX{{%
    \normalfont B\kern-0.5em{\scshape i\kern-0.25em b}\kern-0.8em\TeX}}}

\setcopyright{acmcopyright}
\copyrightyear{2018}
\acmYear{2018}
\acmDOI{XXXXXXX.XXXXXXX}

\acmConference[Conference acronym 'XX]{Make sure to enter the correct
  conference title from your rights confirmation emai}{June 03--05,
  2018}{Woodstock, NY}
\acmPrice{15.00}
\acmISBN{978-1-4503-XXXX-X/18/06}

\usepackage{xcolor}
\usepackage{tabularx}

\begin{document}

\title{Looking at Creative ML Blindspots with a Sociological Lens}

\author{Katharina Burgdorf $^*$}
\affiliation{%
\institution{Manheim University}
 \country{Germany}}

\author{Negar Rostamzadeh $^*$}
\affiliation{%
 \institution{Google Research, Montreal}
 \country{Canada}}

\author{Ramya Srinivasan $^*$}
\affiliation{%
 \institution{Fujitsu Research of America}
\country{USA} }

\author{Jennifer Lena}
\affiliation{%
 \institution{Columbia University}
 \country{USA} }

\def\thefootnote{*}\footnotetext{These authors contributed equally to this work and are listed in alphabetical order}



\begin{CCSXML}
<ccs2012>
<concept>
<concept_id>10003456</concept_id>
<concept_desc>Social and professional topics</concept_desc>
<concept_significance>500</concept_significance>
</concept>
<concept>
<concept_id>10010147.10010178</concept_id>
<concept_desc>Computing methodologies~Artificial intelligence</concept_desc>
<concept_significance>500</concept_significance>
</concept>
</ccs2012>
\end{CCSXML}

\ccsdesc[500]{Social and professional topics}
\ccsdesc[500]{Computing methodologies~Artificial intelligence}

\keywords{social sciences, art, machine learning}


\begin{abstract}
How can researchers from the creative ML/AI community and sociology of culture engage in fruitful collaboration? How do researchers from both fields think (differently) about creativity and the production of creative work? While the ML community considers creativity as a matter of technical expertise and acumen, social scientists have emphasized the role of embeddedness in cultural production. This perspective aims to bridge both disciplines and proposes a conceptual and methodological toolkit for collaboration. We provide a systematic review of recent research in both fields and offer three perspectives around which to structure interdisciplinary research on cultural production: people, processes, and products. We thereby provide necessary grounding work to support multidisciplinary researchers to navigate conceptual and methodological hurdles in their collaboration. Our research will be of interest to ML researchers and sociologists interested in creativity that aim to conduct innovative research by bridging both fields.
\end{abstract}

\maketitle

\section{Introduction}
Recent years have seen a rise of ML applications in creative industries. For example: ML tools are used for analyzing artistic content \cite{shen2019discovery}, for establishing novel methods of creating and re-mixing media \cite{mazzone}, for storytelling \cite{ammanabrolu2020automated}, for creating a variety of animations such as for turning photos into videos \cite{holynski2021}, for various types of recommendation applications \cite{schedl2019,campo2018}, for evaluating creative human performances \cite{kim2019}, and many more.
While the ML community largely characterizes advancements through quantifiable metrics such as efficiency, accuracy, and scalability \cite{birhane2021,hanna2020}, it is important to recognize that from a sociological perspective, creativity is inherently human-centric and collective \cite{godart2020}. Creativity is characterized by heterogeneity of involved people and used materials, both of which can span a wide spectrum--from being extremely organized to wildly chaotic. The ML and sociological perspectives on creativity are thus strikingly different. Researchers argue that overlooking the sociological factors governing creativity can not only hinder the development of meaningful creative ML systems, but also incur a moral cost \cite{loi2020}. This concern has been further aggravated with the advent of deep learning technologies which have enabled end-to-end automatization. Indeed, techniques such as ``\textit{handcrafting}'' or ``\textit{feature engineering}'' are often viewed as mediocrity or weaknesses of ML systems \citep{o2019deep}. On the contrary, creativity demands metaphorical thinking, social interactions, and going beyond extrapolations, which current neural network architectures are far from accomplishing \cite{oleinik2019}.

Creativity in the social sciences has been defined ``as an intentional configuration of material and cultural elements that is unexpected for a given audience" \cite{godart2020}. Creativity stems from a combination of symbols, materials, and ideas through artists that are embedded in a community of fellow artists \cite{becker1982,bourdieu1993}. Every creative act is embedded in larger structures of co-creators that enable or constrain creative configuration. Artists, as well as critics, audiences, and art institutions all shape the creative process and evaluation of creative work. Artists collaborate with other artists, and artists borrow artistic content from other artists. Previous sociological research has studied the antecedents of creativity, highlighting the role of team composition \cite{uzzispiro2005,devann2014}, the specific use of artistic content \cite{lena2004,lena2013}, and arts institutions \cite{lena2019, whitewhite1993}, as well as the consequences of creativity for the evaluation of cultural products by peers, critics, and public audiences \cite{rossman2010,chong2011,cattani2014}. 

To understand potential opportunities and challenges of applying ML tools to creative domains, a profound understanding of the collective nature of creativity is therefore warranted. This paper attempts at bridging the social scientific perspective of creativity with the ML perspective. We discuss where and how creativity is embedded in larger social structures and identify themes and tensions that emerge from the combination of sociological ideas and ML applications. We highlight gaps where more research is needed to understand the interplay between creative work and ML applications. 

To fruitfully bridge a social scientific perspective with a ML perspective, we suggest focusing on issues concerning:
\begin{itemize}
    \item{\it People:} This includes all stakeholders involved in the design and development of creative works. Depending on the use case, there can be different stakeholders. For example, in the context of visual arts, people could imply artists, commissioning agents, auction houses, individual buyers, sellers, etc. In the context of films, directors, producers, cinematographers, and distributors constitute some of the people involved. Within the context of creative ML systems, people could include stakeholders like ML scientists, software engineers, business executives, funding authorities, ML art gallery organizers, digital artists, and so on. 
    \item{\it Process:} The creative process constitutes all aspects right from the conceptualization of the creative idea to the evolution of the creative product. The creative process includes stages such as gathering of resources to facilitate the implementation of the creative idea as well as the actual implementation of the creation to render the creative product. For example, in the context of filmmaking, the process comprises research and material collection, concept and shooting plan development, production, and post-production including editing and sound adjustment.
    In the context of creative ML systems, the process includes problem definition, dataset curation, data annotation/labeling, data processing, model development and validation. Depending on the application, this process could also include a human-in-the-loop for co-creation--e.g., ML tools that support existing artistic practices \cite{kerdreux,davis}.
    \item{\it Products:} Here, by product we mean cultural products which are the outcomes of the creative production process such as a fashion piece, a painting, a film, a song, and so on. In the context of ML creative systems, fashion designs \cite{rostamzadeh2018}, music recommendation systems \cite{schedl2019}, and animations \cite{pat2021} could be some example products. 
\end{itemize}

Focusing on the people-process-product constituting a creative system, we develop a taxonomy to stimulate exchange and further research among the sociology of arts and the ML community. Our study is informed by an multidisciplinary perspective \cite{romm:interdisciplinarity}, comprising of a team whose expertise spans diverse fields such as sociology of music, film making, visual and performing arts, computer vision, machine learning, and applied ethics. To conduct this research, we had to ground discussions across multiple disciplines by mapping and aligning these diverse concepts in being able to surface overlooked issues with respect to the creative ML community. 

We first conducted a systemic review of relevant literature in sociology of culture and creative ML, focusing on the people, process, and products constituting the creative pipelines. We ground our research by analyzing the issues of concern under each of these three overarching themes. The specific issues studied under the three themes are as follows. Within the context of people, we discuss issues concerning team compositions and socio-demographic configurations as these aspects play a pivotal role in the realization of creative ideas. From a sociological perspective as creativity is inherently collective, we focus on the relational dynamics constituting the creative process. Within the context of products, we focus on issues related to assessments of creative products as sociological factors governing evaluation of creative artefacts are seldom considered in ML studies. Across all the three themes, we outline gaps in current creative ML pipelines and suggest pointers for addressing the same. We hope that findings from this study will be useful in the design, development, and deployment of future creative ML systems. 

The reminder of the paper is organized as follows. Each of Section 2-4 first provides an overview of the respective theme (people, process, and product in that order) from a sociological perspective and is then followed by a sub-section that highlights the gaps in existing ML systems. Section 5 discusses reflections from the study. We discuss limitations and future work in Section 6 before concluding in Section 7. 

\section{People: How structural and socio-demographic configurations of people shape creativity}
From a sociological perspective, structural and socio-demographic configurations of people play a pivotal role for the realization of creative ideas. The structural conditions as visible, for example, through artists' collaboration networks, influence how artists organize and accomplish creative work. For example, the authors in \cite{uzzispiro2005} analyze collaboration networks among Broadway musical artists and find that networks that are neither too fragmented nor cohesive, but those that mirror `small-world' properties breed creativity \cite{wattsstrogatz1998}. If networks are too cohesive, this stipulates the free flow of new resources and information. If networks are too fragmented, there is no connection between teams, preventing accessibility of novel information and ideas. In a different study on the gaming industry, the authors in \cite{devann2014} show that creativity emerges from cross-cutting teams of designers that carry different cultural styles. They argue that teams tend to be more creative when their team members tolerate and exploit overlapping group membership and, thus, embrace diverse stylistic approaches \cite{vedresstark2010}. 

Next to team composition, individual network position plays an important role in the creation of cultural products. Sociological research has found that people who occupy brokering position, thus, bridging otherwise disconnected groups, tend to be more creative than people that are constantly embedded within the same group \cite{burt2004}. Brokers tend to have access to different thought communities, ideas, and resources and can recombine elements from different worlds. While structural configurations induce teams and individuals to be more creative, sociologists emphasize that socio-demographic characteristics may hinder the access to structurally favorable teams and positions. In a recent study, \cite{hofstraetal2020} show that underrepresented groups are more likely to produce innovative scientific work, but their work is less valued, and they make less successful careers compared to majority groups. In addition, women are less likely to occupy broker positions that are important for the generation and realization of creative ideas \cite{fangetal2021} and tend to work in closed rather than wide-spanning networks \cite{lutter2015}. In sum, the sociological perspective highlights the role of people's structural position and socio-demographic background for the generation and realization of creative ideas.

\subsection{People in creative ML systems}
%Ramya and Katharina, please check this part.
ML generated artworks and more broadly computational artefacts are generally attributed to digital artists or machine learning researchers. However, it is important to note that ``people'' involved in the creative ML process are not just the individuals who directly create the art piece---there could be multiple factors that can impact individual artists and the reputation associated with their works.

Consider for example, a digital artist who is also an ML expert. Such a person possesses deep knowledge of ML systems and algorithms and can thus leverage their knowledge in the creation process. This expertise sometimes gives them the ability to contribute even on the technical side, which is often extolled and rewarded in ML communities and competitions. These ML artists enjoy the privilege of being known in ML communities, collaborate with other ML researchers and promote their work in the form of ML research papers or open source codes. They are often invited to give talks in ML venues and are surrounded by a clique of ML experts. These privileges give them a platform to present themselves---their story and their art.  Similarly, an artist who is not necessarily a ML expert but who has associations with ML experts may also benefit in multiple ways---from getting access to computational resources to being recognized in the ML community. On the other hand, artists who don't have access to ML community/networks or artists who face geographical barriers in participation, may miss a well-deserved representation and recognition for their work. 

Given this complex interplay between computational artworks with the ML community, multiple aspects come to the discussion. According to World Economic Forum only $22\%$ of ML professionals are women \footnote{\url{https://reports.weforum.org/global-gender-gap-report-2018/assessing-gender-gaps-in-artificial-intelligence/##view/fn-21}}. The same is true with respect to other socio-demographic attributes.
Furthermore, sponsors and their motivation can heavily influence the recognition an artwork receives. Multiple studies have pointed out that the socio-demographic configuration of researchers influences how the work is viewed, cited, or recognized within the research community \cite{hajibabaei2021gender,andersen2019meta,west2019discriminating}. The structural conditions and the influence of team demographics in the ML community, has been a subject of discussion in recent years \cite{west2019discriminating}. Most of the teams and developed datasets in the ML community \cite{rostamzadeh2018,liu2016deepfashion}, have western compositions, or mainly represent North American needs and households \cite{lin2014microsoft}. Art datasets, largely represent cultures and people associated with Western cultures, and are thus not necessarily representative of broader socio-cultural contexts. 

Artists are also likely to be impacted by power, resources, surrounding teams and team compositions, both at an individual level and at a group level \cite{rostamzadeh2018}. For example, a generated art piece submitted to an ML conference art Gallery \cite{rostamzadeh2021ethics} or music festival \cite{huang2020ai}, is typically created by one or more ML artists who have access to certain computational resources (i.e; GPUs, TPUs), connections with a group of ML researchers affiliated with famous universities and/or top tech companies. This may give them an edge in even describing the computational principles behind their art pieces in the form of papers further giving them an edge over other artists who lack such resources and networking.
In sum, it is crucial to note that ML Creativity can be impacted by the nuances and power structures prevalent in the ML community at both individual and group levels. ML Stakeholders represent a small subset of real world social demographics thus leading to not only concentration of power but also suppression of the creative process. Furthermore, due to the emphasis on scale and fast pace research in the ML community, creative advancements are likely to amplify existing biases and ring in new ethical concerns \cite{loi2020}. 

Thus questions concerning team composition and individual structural positions, as well as socio-demographic diversity are relevant to ask for the ML community. How diverse is the ML team, to what extent do artists bridge otherwise disconnected teams to facilitate the flow of information?

\section{Process: How relations shape the creative process}
Sociologists focus on the collective rather than individual underpinnings of the creative process. Previous sociological work has shown that the decisions artists make in the creative process are influenced by relational dynamics that mirror broader social inequalities, group membership, and boundary making. The relational quality of the creative process becomes visible in all steps of the creative process. For example, the successful acquisition of resources artists need to realize their ideas - be it colors and canvases in the visual arts or camera equipment and costumes in film making - heavily depends on artists' previous success and reputation \cite{peters2021}. The successful exhibition, distribution, and promotion of art works depends on relationships artists have to galleries, and other cultural institutions \cite{giuffre1999}. From compiling resources to exhibiting art works, all these steps are shaped by social factors.

A less intuitive but important relational quality of the creative process becomes visible in the use and repetition of artistic content. Pablo Picasso has been quoted as “Lesser artists borrow; great artists steal”. While it is up to debate if he really said this, the phrase points to the fundamental artistic practice of selecting content of other artists' works and incorporating it into one's own work. Even if no collaborators are directly involved, the creative process is inherently collective as creative producers draw from previous art works in their own art works. Artists repeat artistic content that other artists invented, they combine elements from different cultural communities, and borrow stylistic elements from others' previous works. The repetition of artistic content signals group membership and draws boundaries around artistic communities \cite{lena2004, lena2014}.

The repetition of artistic content is, for example, prevalent in the creative process of visual arts, film making, and music. Visual artists pick up artistic content, such as brushstrokes or color palettes, from other artists \cite{becker1982}. Filmmakers include references to earlier films when they adopt camera shots, dialogue snippets, or stills \cite{spitshorvat2014}. Rap musicians borrow content when they use samples of previous songs. For example, \cite{lena2004, lena2013}, show how rap artists incorporate diverse samples into their own songs. The repetition of artistic content not only helps to solve technical problems in the creative process, but gives also rise to new conventions that following artists can build on. While some artists adopt more established content to signal belonging and identity, other artists pursue novel and distinct content to deviate from community conventions.

A major sociological question considering the creative process is how artists navigate the tension between distinctiveness and similarity from other artists and their own previous works. In the context of music production \cite{askin2017} examine how relational proximity or distance of sonic features - ranging from key, tempo, and mode, to energy, or danceability - influences a song's popular success. Songs that sound too similar to other recent songs are less likely to succeed, while songs that reflect optimal differentiation tend to score higher in popular rankings. Cultural embeddedness of stylistic elements shapes the creative process. In the context of fashion, the authors of \cite{godart2019} show how designers adopt stylistic elements, such as colors, fabrics, looks, or patterns, from other designers. They find that stylistic elements are more likely to be adopted when they are structurally embedded among otherwise disconnected elements. The creative process involves not only the decision whether to deviate or build on other artists' styles, but also whether they pursue consistency with regard to their own body of work \cite{wohl2019,coman2020}. 
In sum, sociologists focus on the collective rather than individual underpinnings of the creative process. Recent sociological studies have shown how the decisions in the creative process are shaped by the relational configurations artists and art works are embedded in.

\subsection{Creative ML processes}
 
ML creative processes have at least three important phases, namely, (i) data collection (ii) training the model/developing the algorithms and (iii) validation/evaluation of the developed system. 
Data is an important resource for ML pipelines. Data reflects who is represented and how? Most datasets in ML communities including creative domains, are representations of Western contexts, and reflect stereotypical norms \cite{vries}. For example, MS-COCO \cite{lin2014microsoft} and Flikr 30K datasets \cite{plummer2015flickr30k}, well-recognized captioning datasets only reflects middle-class Western households, and imitates societal gender biases \cite{hendricks2018women,caliskan2017semantics,van2016stereotyping}. The same is applicable to tasks like fashion image retrieval or design generation problems, datasets are usually carrying gender stereotypes, cultural biases and are limited to Western cultures \cite{fashion-gen,liuLQWTcvpr16DeepFashion,wu2021fashion}. Only a handful of datasets try to address this issue by representing broader cultures such as se-Shweshwe Fashion dataset \cite{Malobola2021} and Kaokore, a Japanese art facial expression dataset \cite{tian2020kaokore}. Such non-Western datasets and in turn the ability to create such datasets and use them for creative tasks via ML  could give a platform to unheard voices and address problems related to under-representation. 
For example, some recent generative ML arts, tried to address under-representation leveraging generative art. Jake Elwes presented drag performance using deep fake, and Nouf Aljowaysir by the work Salaf tried to represent her genealogical journey \footnote{https://computervisionart.com/pieces2021/salaf/}. That said, although ML datasets serve as a platform to represent underrepresented voices and stories, restricted access to computational and data resources, and being disconnected from networks of ML artists can hinder the prospects of such under-represented artists. 

ML creative processes are also shaped by relational dynamics of the people involved. Creative ML community often values and prioritizes technical contributions and computational novelty over intrinsic artistic abilities. However, for creative ML tasks, understanding the broader artistic context, culture and history also become important. Furthermore, many creativity-driven ML papers are only presented in creativity workshops and not in the main ML venues, thereby not garnering as much importance as other research topics. In this context, it is also worth noting the influence of citations and group efforts on the recognition received by creative ML research works. 
ML papers are among the most cited papers according to the Natureindex\footnote{https://www.natureindex.com/news-blog/google-scholar-reveals-most-influential-papers-research-citations-twenty-twenty}. The most cited papers often re-use/ manipulate model architectures that are popular (e.g., Generative Adversarial Networks) and may not necessarily explore the possibility of lesser known model architectures which may be still applicable/relevant. 

Furthermore, ML papers and research works are often group efforts. Although most ML conferences have anonymous review process, citations can potentially reveal influential authors associated with a paper, who may have promoted their artworks in different venues. Additionally, workshops in main conferences often invite the most influential and often highly visible researchers to share their findings. Although this may have a positive impact on receiving participants in the workshop, ideas and papers that are not mainstream or those that are not authored by highly influential authors may not get a representation. 

In summary, the creative process, whether in film making, visual arts, or ML-based arts, is fundamentally shaped by relations. Gaining access to resources depends on the connectedness to other members from the field. Furthermore, the creative process also involves more subtle relations when artists repeat content of other artists, an aspect that is largely overlooked in computational measures of creativity. Thus the question of who is connected to whom through what relations is an important consideration for the ML community. One could ask how to avoid the development of highly skewed popularity dynamics, and encourage the collaboration with less visible but highly creative contributors. On the repetition of content side, the ML community may ask how much to value works that are using non popular or non mainstream methods of creation but still bring a niche story.

\section{Products: How creative products are evaluated}

We tend to believe that the assessment of artistic value depends on an artworks' quality, such as technical finesse or novelty. Sociologists, however, have shown that the evaluation of creative work is fundamentally shaped by the type of audience that evaluates - peers, critics, or the public-, artists' embeddedness within the artistic community, as well as historical conditions that shape the criteria for evaluation. The criteria on which these evaluating communities build their assessment on and construct meaning from art works differ significantly. In the sociological literature, peer assessment is often captured through prizes and awards \cite{rossman2010, cattani2014} or through the repetition of other artists' content \cite{lena2004}. Critics communicate their assessment through critical awards or reviews \cite{cattani2014}, and public audiences through sales or box office revenues \cite{lena2013}. 

Depending on the evaluating audience, different criteria can influence the assessment. When peers evaluate their fellow artists' creative work, they tend to favor artworks that adhere to established aesthetic conventions \cite{cattani2014}. Peers build their decisions more on signals of the artist's embeddedness into the community. For example, \cite{rossman2010} find that status and experience of an artist's collaborators significantly influence their chances for an Academy Award nomination \cite{rossman2010}. 

When critics evaluate creative works, they tend to favor those works that deviate from a field's conventions and signal novelty \cite{cattani2014}. Furthermore, critics base their assessment on an artist's socio-demographic status. For example, gender and race fundamentally influence evaluation in critical reviews of books \cite{chong2011,chong2020}. Book critics draw their assessment on racial and ethnic identifiers to claim a novel's authenticity, to assign works into ethnic genres, and to identify talent. 

When public audiences evaluate creative works, they tend to favor works that share similarities with other contemporary works and an artist's previous works but still differentiate to a certain degree. In the case of music, sociological research has shown that public audiences favor songs that are optimally distinctive in their sonic features compared to other contemporary songs \cite{askin2017}. In the case of visual arts, the authors in \cite{sgourevalthuizen2014} show that public audiences only reward stylistic inconsistencies -- differences of an artist's work compared to his or her previous work -- for high-status artists.

It is important to note that many artists seek to create art for art's sake rather than appealing to mass audiences. Community involvement and maintaining an artistic identity are more important than high revenues and popular acclaim \cite{lena2014}. Though many creatives pursue financial disinterestedness, they typically depend on producers, agents, or galleries that hold commercial interests. This conflict of interest may cause artists and producers pulling into different directions throughout the creative process. Producers have a high interest to nudge artists into a direction that ensures revenues, but this may come at the cost of individual artistic expression. 

Another important aspect to raise is that the criteria for evaluations are not fixed but may change depending on the institutional arrangements -- i.e., the role of art museums, critical discourse, and educating institutions -- and the organization of artists within artistic movements \cite{baumann2001, accominotti2009,salganik}. For example, the authors in \cite{whitewhite1993} show that the criteria of evaluation for visual artists changed through the active involvement of the Impressionist art movement. Besides critics, the taste of public audiences may change over time. The author in \cite{lena2019} shows how the appreciation of more diverse art forms in US-American society was facilitated through a shift in historical and institutional arrangements. The social and historical conditions are, thus, important to consider when aiming to understand the underlying drivers of artistic production and evaluation.

\subsection{Assessments of Creative ML Products}
Assessment of creativity is a long studied and central problem in computational creativity \cite{colton}. Researchers opine that there are two main aspects in play in creativity measurement--one that is associated with the creative process and the other that is associated with the creative product which is the outcome of the creative process \cite{colton}. It is argued that this dichotomy between the creative process and the creative output is central to the debates concerning creativity measurement \cite{micro}. Most of the research in computer science focuses on the assessment of creative products though. 

Numerous studies have attempted to evaluate creativity basing the assessments on metrics such as `novelty', `value', and `surprise', to name a few. For example, the authors in \cite{elgammal} quantify creativity of paintings and sculptures using ideas of novelty and `influence' in the context of art networks. Building on this work, the authors in \cite{disha} propose a regression-based learning framework which takes into novelty, influence, value, and `unexpectedness' to evaluate creativity associated with movies. In a similar vein, the author in \cite{mary} proposes to measure creativity of humans and intelligent systems based on novelty, value, and unexpectedness associated with the creative artefacts. In addition to these factors, the authors in \cite{pythagoras} mention `recreational effort' as another factor to consider in evaluation of creative artefacts. Based on the hypothesis that perceptual `ambiguities' play an important role in aesthetic experience, the authors in \cite{wang} present an approach for measuring the perceptual ambiguity of a collection of images. 

Familiar operational definitions of creativity such as those mentioned above only fit well in some cases \cite{boden}. These definitions and metrics of assessment are problematic as sociological factors (e.g., status or embeddedness of the artist as discussed earlier), dynamical factors (e.g., the changing influence of institutional arrangements on evaluations), and historical factors (e.g., pre-existing racial and ethnic inequalities) are seldom considered in the evaluation. ML studies seem to focus mainly on novelty in the assessment of creative work, while sociologists have found that novelty is valued by critics, but often not by other artists or public audiences. Furthermore, computational assessments largely focus on harmony without necessarily considering subjective uncertainties \cite{davis2011}. For example, in a recent study concerning harmony interpretation of popular music, it was shown that annotator subjectivities can play a dominant role \cite{koops2019}. 

Similar to the observation in social science research that artistic inconsistencies are rewarded more for high status artists compared to lesser known ones, it is very much possible that certain generative artworks that are rendered by well known generative artists may receive higher reward in terms of popularity and revenue than other generative artists. In other words, generative artists are probably also embedded within an artistic community that shapes meaning making and assessment of art works. Furthermore, generative artworks created using certain ML models/architectures (e.g., Generative adversarial networks, transformers, etc.) may be perceived to be more favorable and aesthetic as opposed to others, not necessarily because they are actually so but because the underlying models that generated these artworks are popular and considered ``state-of-the-art". With the emergence of ML art galleries and competitions \cite{rostamzadeh2021ethics}, factors outlined earlier such as social networks and marketability can also influence assessment. 

In summary, different metrics matter differently based on the broader socio-cultural-historical contexts concerning the creative product. Furthermore, evaluations are also dependent on the human evaluator, respectively the evaluating community, an aspect that is largely overlooked in computational measures of creativity. Thus questions such as who is seeking the assessment of the ML creative product, what is the rationale behind the creation of the product, for what reason is the assessment being made, when is the assessment being made become important considerations in computational creativity assessments.

These aspects discussed are applicable both in sociological contexts and in ML contexts. Apart from structural team compositions and social demographics of the people involved in creative pipelines, community protection (i.e. safeguarding the interests of creative people in the context of technologies like ML, ensuring that underrepresented artists are not sidelined, etc.) is also important. In the context of creative process and creative products, we largely focused on relational dynamics underpinning art creation and assessments of products. Another important aspect concerns meaning making of artistic process and artistic products---understanding the larger socio-cultural contexts associated with creative processes and products, to be able to interpret and appreciate the perspective of the artist, and so on. And finally, regulatory norms such as copyright issues, ownership rights in the context of ML created artefacts, and accountability become relevant in the context of creative products.  

\section{Discussion}

Here, we reflect on some open questions that arise at the intersection of sociology, art, and ML within the context of the three themes discussed above, namely, people, process, and products. 

As outlined in previous sections, creativity is inherently collective, shaped by the values, beliefs, preferences, and experiences of a diverse set of people. Two fundamental questions arise in this regard ---To what extent can algorithms `imitate' the collective nature of creative production? Can algorithms `imitate' diversity in team composition or socio-demographic characteristics? Researchers argue that creative ML technologies often suffer from crucial aspects of human cognition such as intuition, abstraction, analogy making, and common sense \cite{grba2022}. It has been argued that although ML technologies offer a generous space for conceptual, as well as formal, methodological, and aesthetic experimentation, they are largely uneven--not representative of the diversity that exists in the real world \cite{grba2022}. Biased and constrained contextual awareness coupled with lack of broader socio-cultural and art-historical knowledge can come in the way of replicating the collective nature of creativity \cite{grba2022}. 

With respect to the creative process, questions related to computational modeling styles of artists call for examination. The author in \cite{grba2022} argues that ML technologies work on the assumption that all relevant features can be found within the training data; but this is problematic in real life scenarios, especially in experiencing and constructing meaning of art where the overall context, audience’s knowledge, and expectations are essential. While ML algorithms may be adept at recognizing low level features such as color, brushstrokes, frequency-related information, they are far from modeling artists' styles, and how people perceive and make meaning of an artwork \cite{forbes2020}. In a similar vein, the authors in \cite{uchino2021} argue that it is not possible to computationally model an artist's style owing to the influence of several unobservable cognitive factors such as emotions and beliefs on the artist's style. To illustrate the same, the authors discuss the shortcomings associated with landscapes claimed to be rendered in the style of `Impressionism art movement' using CycleGAN, a popular ML model \cite{cyclegan}. The authors demonstrate that subtle aspects of Impressionism such as spontaneous and accurate depiction of light with its changing colors are not captured in the generated artworks and further state that the generated artworks do not do justice to the cognitive abilities of the artist. The authors add that in computationally modeling an artist's style, artists could be even stereotyped based on narrow metrics like color and brushstrokes. 

Related to the concern associated with computational modeling of artists' styles is the issue of reinforcement of human biases by ML algorithms. For example, the authors in \cite{wang2019} ask if a logistic classifier can detect a female style in music based on various features of songs, ranging from sonic features to artist's tagging by listeners, and collaboration networks. Using crowdsourced and algorithmically augmented data they identify the existence of a female style in music which raises questions on potential reproduction of gender biases. If gender-related features are, for example, highly correlated with revenues, inequalities in the music industry may perpetuate. It also raises the question on the quality of crowdsourced training data. If user entries are shaped by gender biases, they may provide more complete information for female as compared to male artists. Another point to elaborate on is-- to what extent a binary understanding of socially constructed categories, such as gender, may perpetuate biases and limit the accuracy of ML models. Yet, several works leverage such bias reinforcing correlative features in creative decision making and predictions \cite{lash2016,kim2019}

ML based creative products have gained enormous popularity in recent years. Amidst this popularity surge, several issues of concern have emerged across different creative ML products such as recommendation systems, generative artworks, fashion designs and more. For example, in a recent work \cite{srinivasan2021}, biases associated with generative artworks were examined through case studies. Delving deeper into fundamental questions concerning ethical relevance of creative ML systems, the authors in \cite{loi2020} emphasize that creativity in very broad philosophical sense, encompasses natural, existential, and social creative processes, such as natural evolution and entrepreneurship, and state that understanding creativity in this way is instrumental for advancing human-well being in the long term. Although works like \cite{mcgregor2020} investigate how computational creativity can be explored to address certain philosophical questions, the authors emphasize that any `computational solution' is likely to open the door for other philosophical debates.  
Another widely debated issue relates to the ideal behind creation of creative ML products. With rapid automization and scaling, there is increased adoption of creative ML products. These are primarily governed by commercial interests such as increased revenue and seldom care about purpose and meaning of artworks. As the authors in \cite{daniele2019} note, art initiates a conversation with the public. In fact, researchers state that artworks enable one to understand what ought to do and what not to do--they impart ``moral knowledge" \cite{young,novitz}. They help one to understand different world views, thereby engendering empathy \cite{leroi}. It is important to recognize that creative products are not merely aesthetic objects but are symbolic of human-emotions, ideals, aspirations, and their relations. 

\section{Limitations and Future Work}
In this work, we attempted to propose a sociological lens on creativity to elucidate potential blind spots in the creative ML community. While we aimed at providing a broad overview on the social science perspective of creativity, our study does not necessarily provide a comprehensive coverage of all important questions at the intersection of ML, art, and society. Future research could elaborate more on aspects of community protection in the context of creative ML. How do artists deal with the fact that advanced technical tools may claim parts of their jurisdiction \cite{sachs2020}? Another important aspect is how artists and audiences deal with issues of lacking artistic legitimation for digital artworks outside of the ML community. Consumers of culture (audiences but also “gatekeepers”) often “make sense” of art pieces by “making sense” of whether the artist “makes sense as an artist.” If generative artists are self-taught and not trained at renowned art institutions, if their work reflects no previous tradition -- simply because this art form is novel --, if art works cannot be assigned to an established style or individual authorship, how can peers and audiences still construct meaning and further develop this novel art form? \cite{fine2003, baumann2007}. Other questions social scientists, ML scientists, and legal scholars may collaboratively work on could address the aspect of copyright and royalties in generative artworks \cite{robertson2022}.

\section{Conclusions}
In this paper we aimed to build a bridge between the social sciences and the creative ML community. We explored how conceptual and methodological approaches from the social sciences can inform the ML community in the design, development, and deployment of creative ML systems. We illustrated the collective nature of creativity and have proposed a taxonomy that guides the study of creativity through a sociological lens.  The taxonomy focuses on people, processes, and products as fruitful perspectives for the study of creative ML. On the aspect of people, we showed that the generation and realization of creative ideas -- whether in the ML art community or in other art worlds -- is shaped by people’s structural position in art networks and artists' socio-demographic background. On the aspect of processes, we highlighted how the creative process -- from compiling resources, to creating art, to exhibiting works -- is shaped by relational configurations. On the aspect of products, we shed light on the factors that influence the assessment of cultural products. We hope that our paper provides a starting point for bridging social science and ML approaches to creativity and to contribute to further discussions on topics at the intersection of arts, AI, and society.

\bibliographystyle{ACM-Reference-Format}
\bibliography{sample-base}

\end{document}